**Tailoring epitaxial growth and magnetism in La$_{1-x}$Sr$_x$MnO$_3$ / SrTiO$_3$ heterostructures via temperature-driven defect engineering**

Alan Molinari,[1,2,*] Saleh Gorji,[2,3] Jan Michalička,[4] Christian Kübel,[2,3,5] Horst Hahn,[2,3] Robert Kruk[2,§]

[1]IBM Research Europe – Zurich, Säumerstrasse 4, 8803 Rüschlikon, Switzerland

[2]Institute of Nanotechnology, Karlsruhe Institute of Technology, Hermann-von-Helmholtz-Platz 1, 76344 Eggenstein-Leopoldshafen, Germany

[3]KIT-TUD-Joint Research Laboratory Nanomaterials, Technical University Darmstadt, Otto-Berndt-Str. 3, 64287 Darmstadt, Germany

[4]CEITEC Nano Research Infrastructure, Brno University of Technology, Purkyňova 123, 612 00 Brno, Czech Republic

[5]Karlsruhe Nano Micro Facility, Karlsruhe Institute of Technology, Hermann-von-Helmholtz-Platz 1, 76344 Eggenstein-Leopoldshafen, Germany

*E-mail: molinarialan86@gmail.com
§E-mail: robert.kruk@kit.edu

**ABSTRACT**

Among the class of strongly-correlated oxides, $La_{1-x}Sr_xMnO_3$ – a half metallic ferromagnet with a Curie temperature above room temperature – has sparked a huge interest as a functional building block for memory storage and spintronic applications. In this respect, defect engineering has been in the focus of a long-standing quest for fabricating LSMO thin films with highest quality in terms of both structural and magnetic properties. Here, we discuss the correlation between structural defects, such as oxygen vacancies and impurity islands, and magnetism in $La_{0.74}Sr_{0.26}MnO_3/SrTiO_3$ (LSMO/STO) epitaxial heterostructures by systematic control of the growth temperature and post-deposition annealing conditions.

Upon increasing the growth temperature within the 500 – 700 ˚C range, the epitaxial LSMO films experience a progressive improvement in oxygen stoichiometry, leading to enhanced magnetic characteristics. Concurrently, however, the use of a high growth temperature triggers the diffusion of impurities from the bulk of STO, which cause the creation of off-stoichiometric, dendritic-like $SrMoO_x$ islands at the film/substrate interface. As a valuable workaround, post-deposition annealing of the LSMO films grown at a relatively-low temperature of about 500 ˚C permits to obtain high-quality epitaxy, atomically-flat surface as well as a sharp magnetic transition above room temperature and robust ferromagnetism. Furthermore, under such optimized fabrication conditions possible scenarios for the formation of the magnetic dead layer as a function of LSMO film thickness are discussed. Our findings offer effective routes to finely tailor the complex interplay between structural and magnetic properties of LSMO thin films via temperature-controlled defect engineering.

**INTRODUCTION**

La$_{1-x}$Sr$_x$MnO$_3$ (LSMO) is a mixed-valence manganese oxide known for exhibiting a robust inherent coupling between lattice, charge and spin degrees of freedom.[1–6] Such intrinsic correlation lies at the origin of an intriguing interplay between magnetic and electric properties, which is manifested in a variety of physical phenomena, including colossal magnetoresistance, half metallicity, metal-to-insulator and para-to-ferromagnetic transitions in proximity to room temperature. This unique combination of physical properties has denoted LSMO as one of the most attractive constitutive elements of novel functional devices such as spin valves,[7–9] magnetic field sensors,[10] magnetoelectric[11–17] and memristive memories[18–21].

Aiming at harnessing the magnetic properties of LSMO for potential technological applications, particular efforts have been devoted to explore the complex relation between microstructure and magnetism,[22–31] particularly for the compound with a strontium content $x \approx 0.3$, for it displays the highest Curie temperature of the whole class of strongly-correlated manganese oxides. In first approximation, magnetism in LSMO can be explained in terms of the competition between double- and superexchange interactions,[1,32] which are strictly dependent on the bond angle and length occurring along Mn-O-Mn chains, and on the oxidation state of the Mn ions (i.e., either 3$^+$ or 4$^+$).

A well-established strategy to manipulate magnetism in LSMO is to vary the degree of distortion of its perovskite structure, and thus bond lengths and angles, by adjusting the magnitude and kind (compressive vs. tensile) of epitaxial strain using lattice-compatible single-crystal substrates, e.g. SrTiO$_3$ (STO), LaAlO$_3$ (LAO), (La$_{0.3}$Sr$_{0.7}$)(Al$_{0.65}$Ta$_{0.35}$)O$_3$ (LSAT), NdGaO$_3$ (NGO) and MgO.[22,23,33–35]

Apart from epitaxial strain, the magnetic exchange interactions in LSMO can be substantially affected by the presence of structural defects, since they are responsible for altering the perovskite structure of LSMO and its local chemical environment.[23,24,30,31,36,37] For instance, considering the local microstructure of the LSMO lattice, the aftermath of a single oxygen vacancy is (i) to disrupt an exchange interaction and so also the hopping of charge carriers along a Mn-O-Mn chain, (ii) to modify the local oxidation state of Mn ions due to doping with two holes, and (iii) to distort the bond lengths and angles of the oxygen octahedron surrounding a Mn ion, thus modifying the respective crystal field.[1,32] Structural defects in LSMO are also considered as one of the principal causes for the occurrence of the so-called magnetic dead layer (MDL),[37–40] which is commonly invoked to explain the lower values of Curie temperature and magnetic moment attained in thin and ultrathin films as compared to bulk LSMO.

From another perspective, structural defects are also of great importance when the purpose is to combine LSMO with other functional materials. Indeed, in several circumstances LSMO is implemented as a magnetic and/or conducting seed layer in combination with dielectric or ferroelectric materials, such as BFO, BTO or PZT.[16,20,41,42] Therefore, an atomically smooth surface is a prerequisite for achieving high quality epitaxial heterostructures. The presence of a large interface roughness, for instance due to non-stoichiometric islands on the LSMO surface, can jeopardize the overall quality of the heterostructure and produce negative side effects, such as reduced interfacial coupling, leakage current or inefficient spin/charge injection.[41,43,44]

For these reasons, defect engineering based on the judicious control of the fabrication conditions represents a pivotal aspect for mitigating the detrimental effects of defects on the structural and magnetic properties of LSMO thin films. Owing to the vast space of growth parameters and their non-trivial interplay, identifying the optimum experimental conditions to attain state-of-the-art LSMO films can pose a formidable task. A well-known approach is to study the influence of various oxygen pressures during film growth, particularly with the aim of optimizing the oxygen stoichiometry.[23,24,30,31,45] Another crucial parameter, whose relation with the used oxygen pressure is often not investigated in depth, is represented by the growth temperature $T_{gr}$, which regulates several aspects of defect engineering, including the film oxidation rate, the migration and nucleation of elemental species on the sample surface and the film/substrate elemental interdiffusion.[46,47] Besides, following film growth, post-deposition annealing under conditions of high temperature and oxygen pressure is an additional tool, which allows to optimize the oxygen stoichiometry and to promote surface reconstruction.[26]

In this work, we examine the role of $T_{gr}$ and post-deposition annealing to finely tailor the structural and magnetic properties of LSMO/STO epitaxial heterostructures. The use of a high $T_{gr}$ of about 700 °C fosters film oxidation, which is reflected by improved magnetic properties; nonetheless, it also provokes the formation of islands at the LSMO/STO interface due to impurities diffusion from bulk STO. A compromise to achieve optimum structural and magnetic characteristics is found upon employing a relatively-low $T_{gr} \approx 500$ °C followed by post-deposition annealing treatment. Eventually, we propose a critical discussion on the formation of the MDL in LSMO as a function of film thickness in case of films grown under optimized conditions.

**EXPERIMENTAL DETAILS**

LSMO thin films were deposited onto (001)-oriented STO substrates by rf-magnetron sputtering (Createc Fischer GmbH). An unconventional characteristic of the used sputtering chamber is the large target-substrate separation of about 30 cm, which permits to find an optimum balance between film homogeneity, deposition rate and suppression of oxygen ion bombardment.[23] In the present study both epi-polished and chemically-treated (see Supplementary Information) STO substrates have been put to test; in both cases LSMO films with similar quality were obtained.

LSMO films with a thickness in the range of 3 – 15 nm were deposited from a commercial sputtering target with chemical composition $La_{0.65}Sr_{0.35}MnO_3$, using a fixed pressure of 0.018 mbar in a 3/2 mixture of $Ar/O_2$, which corresponds to an oxygen partial pressure of 0.007 mbar. The sputtering power was set to 4.4 W/cm², which led to a deposition rate of about 0.044 Å/s. The growth temperature was varied between 500 – 700 ˚C. After the growth process, the LSMO/STO samples were cooled down to room temperature at a rate of 10 K/min in pure oxygen atmosphere of 0.08 mbar. Post-deposition annealing was carried out by exposing the as-grown LSMO films to 900 ˚C for 1 hour in air inside a tube furnace.

The LSMO films have been comprehensively characterized with a variety of experimental techniques. Rutherford Backscattering Spectroscopy was carried out on a LSMO film grown on MgO substrate in order to avoid the overlap of Sr signals and revealed a strontium content of about Sr = 0.26.[48] The offset between LSMO target and film concentration is ascribed to the different mean free path of elemental La and Sr species during the sputtering process. For the resulting LSMO stoichiometry, the calculated interfacial misfit with STO substrate is $\varepsilon = (a_{STO} - a_{LSMO})/a_{LSMO} \approx 0.45\%$, where $a_{LSMO} = 3.888$ Å and $a_{STO} = 3.905$ Å are respectively the lattice constants of bulk pseudocubic LSMO and bulk STO. X-ray reflectometry was used to measure the LSMO film thickness, whereas high-resolution X-ray diffraction (HRXRD) in $\vartheta$-$2\vartheta$ geometry was carried out to determine the structural quality of the LSMO films along the growth direction. The surface roughness of the LSMO thin films was analyzed via tapping-mode atomic force microscopy (AFM). Local information about the film microstructure and composition was evaluated by scanning transmission electron microscopy (STEM) using a FEI Themis 60-300 cubed equipped with a Super-X energy-dispersive X-ray (EDX) detector. The representative LSMO cross sections were prepared by focused ion beam (FIB) milling using a FEI Strata 400S.

The magnetic properties of LSMO films were investigated by superconducting quantum interference device (SQUID) magnetometry. The magnetic transition temperature of LSMO films was analyzed by carrying out field-cooled measurements in the 2 – 350 K temperature range at a fixed magnetic field of 100 Oe applied parallel to the in-plane film direction. Magnetization saturation and coercivity have been evaluated by performing magnetic hysteresis loops up to a magnetic field of ± 9 T at a temperature of 10 K.

**RESULTS AND DISCUSSION**

The structural investigation by means of HRXRD reveals that all LSMO films with a thickness of about 13 nm deposited onto STO substrate in the 500 – 700 ˚C temperature range are epitaxial and <001>–oriented along the out-of-plane direction (see Fig. 1). The presence of pronounced Laue oscillations in proximity of the (002) LSMO reflection indicates the coherent stacking of the LSMO unit cells along the growth direction, thus confirming a high degree of epitaxial quality.

The main influence of $T_{gr}$ is to induce a systematic shift of the HRXRD peaks of LSMO. For $T_{gr} \approx$ 500 – 550 ˚C the LSMO and STO peaks almost overlap, whereas upon increasing the $T_{gr}$ up to 700 ˚C the LSMO reflections systematically shift towards higher angles (see Fig. 1 top), corresponding to a decrease in the out-of-plane lattice parameter of LSMO from about 3.90 Å to 3.86 Å. It is established that the LSMO unit cell undergoes an expansion along the c-axis in the presence of oxygen-deficiencies.[23,49] Therefore, the progressive shrinkage of the LSMO unit cell at higher values of $T_{gr}$ provides a clear indication of the improvement of the oxygen stoichiometry in the LSMO films. This observation is confirmed by the results of the in-depth magnetic characterization described in the following.

The abundant presence of oxygen deficient sites at low deposition temperatures is further corroborated by analyzing the effect of post-deposition annealing at 900 °C for 1 hour in air on the as-grown samples. In this case, the HRXRD peaks of all LSMO films, which still feature neat Laue fringes, shift towards a common $2\vartheta$ value, corresponding to an out-of-plane lattice parameter of about 3.86 Å (see Fig. 1 bottom). This outcome strongly indicates that a post-deposition annealing treatment permits to adjust the oxygen stoichiometry of all LSMO films in a comparable manner regardless of the initial $T_{gr}$.

The crystalline quality of the LSMO films was also inspected by studying the rocking curves ($\omega$-scans) of the LSMO (002) reflections (see Supplementary Fig. 1). The full-width at half maximum (FWHM) of the $\omega$-scans is 0.1° and 0.06° for the annealed films grown at 540 °C and 700 °C, which are among the best values reported in the literature.[23,49,50] The low values for the peak width of the rocking curves suggest only a minor misalignment of the LSMO crystalline planes with respect to the STO substrate orientation. Information about the in-plane epitaxial relation between film and substrate have been accessed via two-dimensional reciprocal space maps of asymmetric (204) XRD reflections, which reveal a nearly ideal, fully strained growth of LSMO onto STO (see Supplementary Fig. 1). Besides, after carrying out a $\varphi$-scan, it is concluded that the LSMO unit cells arrange in a coherent tetragonal stack onto STO for the conditions of four-fold symmetry are fulfilled (see Supplementary Fig. 1).

The surface morphology of the post-annealed LSMO films grown at various temperatures was investigated by atomic force microscopy (see Fig. 2). For values of $T_{gr}$ < 600 °C the LSMO surface is characterized by atomically flat terraces with a height of one or half unit cell steps over large areas of several µm$^2$ and a corresponding root mean square (RMS) roughness lower than 0.2 nm. The presence of flat terraces is the result of surface reconstruction after post-deposition annealing, since the as-grown films are atomically smooth, but do not have step-like features (see Supplementary Fig. 2). Notably, above a critical temperature of $T_{gr} \approx$ 600 °C, islands are formed on the LSMO surface, which cause an increased surface roughness, though flat terraces are still visible. An increased surface roughness with increasing $T_{gr}$ has also been observed in pulsed laser deposited LSMO films.[51] It is worth to note that at $T_{gr} \approx$ 650 °C, the islands have an average diameter of about 100 – 200 nm and are randomly distributed on the LSMO surface. Differently, at $T_{gr} \approx$ 700 °C the LSMO surface presents large dendritic-like islands with a lateral size of about 1 µm separated by wide portions of the LSMO surface with atomically-smooth, step-like terraces. This behavior suggests that $T_{gr}$ acts as a driving force to promote initially islands formation (for $T_{gr}$ > 600 °C) and then islands coalescence (for $T_{gr}$ > 650 °C).

Apparently, one may argue that such islands are composed of chemically-spurious species, that diffuse and merge at the top surface of the LSMO film. Previous studies already proposed that non-stoichiometric islands form on the uppermost layers of LSMO due to adatom inhomogeneities that do not overcome a critical island size[44] or due to strontium segregation driven by a buildup of oxygen vacancy-strontium interactions near the LSMO surface regions.[52]

However, a local inspection of the structural and chemical features by a combination of STEM and EDX methods of an island formed at $T_{gr} \approx 700$ °C unveils a different scenario for the process of island formation in our LSMO/STO heterostructures (see Fig. 3).

On the one hand, the region of the heterostructure far from the surface island present a smooth top surface and cube-on-cube epitaxy of LSMO onto STO (see right side of Fig. 3), thus corroborating the high crystalline quality and fully-strained growth already identified by the HRXRD study. On the other hand, STEM analysis in the proximity of a surface island clearly evidences the presence of a grain formed near to the film/substrate interface rather than on top of the LSMO film (see left side of Fig. 3). A closer inspection of the film/substrate interface reveals that the spurious grain is not directly in contact with the STO substrate, but lies onto a few unit cells of LSMO. Most strikingly, despite the interfacial grain the heterostructure maintains a coherent epitaxial growth with the underlying STO substrate, including the LSMO film on top. Concerning the chemical composition, EDX analysis reveals that the interfacial grain is composed of a $SrMoO_x$ phase with a nearly 1-to-1 Sr/Mo ratio, whereas the other regions of the heterostructure correspond to LSMO and STO.

The combination of STEM and EDX surveys indicate that the nucleation of SrMoOx grains occurs during the growth process of LSMO film as a consequence of temperature-driven diffusion of impurities from bulk STO. We note that the role of defect chemistry and the importance of background impurities in undoped STO are still under intensive evaluation.[53–55] In a plausible scenario, during the initial stages of film deposition, the interfacial islands are small enough to permit the formation of a few coherent and continuous layers of LSMO onto STO substrate. As the deposition advances, more and more impurities diffuse from the STO substrate and provoke the expansion of the interfacial grains, which are progressively covered by the growing LSMO film. Concerning the evolution of the spurious grains as a function of different growth temperatures, we note from the AFM analysis that they do not have a preferential spatial distribution up to $T_{gr} \approx 650$ °C (see Fig. 2c). However, higher temperatures of $T_{gr} \approx 700$ °C (see Fig. 2d) promote a higher mobility of Sr and Mo species, thus resulting in the nucleation and coalescence of extensive $SrMoO_x$ grains at the LSMO/STO interface.

The fact that $T_{gr}$ acts as a driving force for the formation of the interfacial grains is further proven by the following test: after simply exposing an original STO substrate to a high temperature of about 700 °C in the reducing atmosphere of the sputtering chamber (without LSMO deposition), dendritic-like islands form on its surface (see Supplementary Fig. 3). Furthermore, we note that previous works already reported on the creation of Sr-rich islands on the STO surface due to Sr-diffusion from the bulk under high temperature conditions.[56–58] Additional results of STEM and EDX analyses of the interfacial grains formed in LSMO/STO heterostructures are provided in Supplementary Figures 4 – 8.

We shall now discuss the influence of $T_{gr}$ and post-deposition annealing on the magnetic properties of epitaxial LSMO thin films with a thickness of about 13 nm (Fig. 4). The magnetic field-cooled $M(T)$ curves of the as-grown LSMO films display a progressive increase in Curie temperature $T_c$ from 110 K to 270 K upon increasing $T_{gr}$ from 500 to 700 °C (Fig. 4a top); the derivatives of the $M(T)$ curves present a FWHM of about 48 K, thus indicating a rather smeared out para-to-ferromagnetic transition (Fig. 4a bottom). Concurrently, the saturation magnetization $M_s$ reveals a systematic increase from 1.4 $\mu_B$/u.c. to 3.3 $\mu_B$/u.c (Fig. 4b). The improvement in $T_c$ and $M_s$ at higher growth temperatures supports the idea of a strong reduction of the amount of oxygen vacancies in LSMO, thus corroborating the results of the XRD analysis discussed above.

After carrying out post-deposition annealing on the as-grown LSMO films, all samples exhibit similar magnetic characteristics with an increased $T_c \approx 320$ K, a sharper magnetic transition (FWHM of d$M$/d$T \approx 25$ K) and $M_s$ above 3 $\mu_B$/u.c. Notably, after post-deposition annealing the LSMO film grown at the lowest temperature of 500 °C undergoes a massive jump in $T_c$ of about 200 K and the value of $M_s$ more than doubles. Although all samples were exposed to the same post-deposition annealing protocol, the LSMO film grown at 700 °C reaches the highest $M_s$ of about 3.5 $\mu_B$/u.c.. On the one hand this is possibly related to a more uniform and complete oxidation achieved throughout the whole LSMO film thickness at a high $T_{gr}$ of 700 °C ; on the other hand the use of a high $T_{gr}$ of 700 °C may favor the formation of straightest Mn-O-Mn bonds, as supported by the sharpest values of rocking curve obtained in the XRD analysis (see Supplementary Fig. 1). Besides, it can also be concluded that the SrMoO$_x$ islands formed at the film/substrate interface due to temperature-driven bulk diffusion from STO do not dramatically affect magnetism in LSMO, if not on a local scale.

It is worth noticing that all LSMO films present a lower value of $M_S$ as compared to bulk LSMO ($M_{Bulk} = (4-x)\mu_B \approx 3.74\,\mu_B/u.c.$ for a Sr content of $x$ = 0.26). In literature this behavior is generally explained in terms of a magnetic dead layer (MDL) formed in LSMO thin films due to a variety of possible factors, comprising magneto-electronic phase separation,[38,59,60] substrate/film elemental interdiffusion,[37,61,62] oxygen vacancies,[40,45,63] or preferential occupation of the out-of-plane Mn 3$d$ $e_g$ (3$z^2 - r^2$) at the expense of the in-plane 3$d$ $e_g$ ($x^2 - y^2$) band.[64,65]

In order to infer more details about the origin of the MDL, we have compared the $M(T)$ curves and the hysteresis loops (see Fig. ) of LSMO films with various thicknesses grown under the same optimized conditions, i.e., deposition at $T_{gr} \approx 540$ °C followed by post-deposition annealing treatment. Looking at three representative LSMO film thicknesses, 12, 4.5, 3 nm, a systematic decrease in $T_c \approx 311, 284, 222$ K, and $M_S \approx 2.7, 1.9, 1.6$ $\mu_B$/u.c., as well as an increase in $H_C \approx 30, 52, 96$ Oe have been observed. To a first approximation, one can estimate the thickness of the MDL formed in each LSMO film by considering the deviation of $M_S$ with respect to the bulk value $M_{Bulk} \approx 3.74 \ \mu_B$. The analysis points out an MDL thickness of about 3.3 nm, 2.1 nm and 1.7 nm for the 12, 4.5 and 3 nm LSMO films. We also note that the absolute value of the MDL is not constant, but becomes smaller upon decreasing the LSMO thickness. Besides, the relative portion of the LSMO film volume being affected by the MDL increases noticeably from 27,5 % to 46 % and 70% for the 12, 4.5 and 3 nm LSMO films. This suggests that a variety of factors may concurrently contribute to the formation of the MDL as a function of film thickness.

First, we observe that all LSMO films with a thickness below 12 nm present a pronounced kink with an upturn in the $M(T)$ curves at about 105 K, which corresponds to the antiferrodistortive phase transition of the STO substrate.[66] The robust magnetoelastic coupling originated at the film/substrate interface supports the conclusion of fully-strained epitaxial growth of LSMO. Therefore, the reduced degree of strain relaxation within the LSMO films and the high-quality of the fabricated heterostructures with cube-on-cube epitaxy suggest that possible effects related to the distortion of the perovskite unit cell of LSMO, e.g. affecting Mn-O-Mn bond lengths and angles, shall not predominantly contribute to the formation of the MDL in our samples.

One possible aspect to consider is that during post-deposition annealing treatment the initial oxidation front formed at the LSMO surface may hamper the diffusion of further oxygen species towards the vacancy sites located deeper in the LSMO film. According to this scenario, ultrathin LSMO films undergo a more uniform and complete oxidation as compared to thicker LSMO films, thus explaining the smaller value of MDL encountered upon decreasing film thickness. Concurrently, however, ultrathin LSMO films may be more sensitive to local interfacial effects, such as film/substrate elemental interdiffusion and the nature of the termination layers on the LSMO film surface, such as missing Mn-O-Mn bonds or modification of the oxidation state of Mn ions. Although interfacial effects may play a role in the MDL formation, they are expected to mostly affect the value of magnetization $M_s$ rather than the magnetic transition temperature $T_c$ of the whole LSMO film. Indeed, it is anticipated that at least a few LSMO unit cells shall remain unaffected by interfacial effects and contribute to the onset of ferromagnetism via conventional double exchange interactions even in case of an ultrathin LSMO film with a thickness of 3 nm. Nonetheless, the fact that $T_c$ does not seem to be simply related to a smeared out para-to-ferromagnetic transition, but rather to a clear shift towards lower temperatures as the thickness is progressively decreased, strongly suggests that additional factors must be taken into account. As supported by previous studies,[38,67] the decrease in both $M_s$ and $T_c$ observed in thin and ultrathin LSMO films is compatible with the phenomenon of magneto-electronic phase separation. In this scenario, intrinsic inhomogeneities within the LSMO lattice cause the coexistence of ferromagnetic/metallic and non-ferromagnetic/insulating phases. It was proposed that such instabilities can be promoted by the reduction of charge carriers at the interface between ferromagnetic/metallic (such as LSMO) and insulating (such as STO) regions.[59] The occurrence of magneto-electronic phase separation is also backed by the results of coupling coefficients obtained in electric-field tuning experiments of similar LSMO films gated with ferroelectrics[39] and ionic liquids.[15,48] In the light of these observations, we conclude that magneto-electronic phase separation is plausibly a major contributor for the formation of the MDL in thin and ultrathin LSMO films grown onto STO substrate.

**CONCLUSIONS**

In this work, the structural and magnetic properties of LSMO thin films grown onto (001)-oriented STO substrates have been investigated as a function of $T_{gr}$ (500 – 700 ˚C) and post-deposition annealing. The as-grown films deposited at $T_{gr}$ < 540 ˚C are atomically flat, however the abundant presence of oxygen deficiencies causes a reduced $T_C$ and $M_S$. The increase of the $T_{gr}$ up to 700 ˚C leads to an enhanced oxygen incorporation, and thus also to a systematic improvement of the magnetic properties. In spite of that, at the same time the use of a high $T_{gr}$ fosters the creation of spurious, dendritic-like islands at the film/substrate interface owing to the diffusion of bulk Mo and Sr impurities from the STO substrate. The preservation of the LSMO film epitaxy as well as the localized character of the $SrMoO_x$ grains suggest that such kind of defects shall not affect the magnetic properties of LSMO on a macroscopic scale. Still, they may cause deleterious effects when the purpose is to realize more complex heterostructures relying on the use of LSMO as an atomically-flat underlayer.

High-temperature post-deposition annealing of the as-grown films at 900 ˚C for 1 h in air has the twofold effect of promoting surface reconstruction and improving the oxygen stoichiometry. Combining the results of XRD, AFM and SQUID analyses, the best compromise in terms of crystalline quality, surface smoothness and robustness of the magnetic properties is attained when employing a relatively-low $T_{gr} \approx 500 - 540$ °C followed by post-deposition annealing. Under such optimized growth conditions, various insights are inferred on possible factors contributing to the formation of the MDL in thin and ultrathin LSMO films, including interfacial effects, gradient in film oxidation and magneto-electronic phase separation.

In conclusion, our study evaluates functional strategies for manipulating the structural and magnetic properties of epitaxial LSMO thin films via temperature-driven defect engineering, hence promoting their exploitation as effective building blocks for magnetoelectric and spintronic applications.

**Acknowledgements**

The authors acknowledge Philipp Leufke and Ralf Witte for support in the fabrication and characterization of LSMO films. This project has received funding from the European Union's Horizon 2020 research and innovation programme under the Marie Sklodowska-Curie grant agreement No 898113 (InNaTo) and from the Deutsche Forschungsgemeinschaft (DFG) under project numbers LE 3519/1-2, and MU 333/2-2 and as part of the DFG Research Group 2093 (Memristive elements for neural systems) under project CH 1492/1-1. Saleh Gorji acknowledges his PhD scholarship by the DAAD. The authors acknowledge the support of the Karlsruhe Nano Micro Facility (KNMF) for the usage of various experimental facilities and CzechNanoLab Research Infrastructure supported by MEYS CR (LM2018110) for TEM measurements.

**References**

[1] E. Dagotto, T. Hotta, and A. Moreo, Phys. Rep. **344**, 1 (2001).

[2] Y. Tokura and Y. Tomioka, J. Magn. Magn. Mater. **200**, 1 (1999).

[3] K. Dörr, J. Phys. D. Appl. Phys. **39**, (2006).

[4] O. Chmaissem, B. Dabrowski, S. Kolesnik, J. Mais, J. Jorgensen, and S. Short, Phys. Rev. B **67**, 1 (2003).

[5] B. Dabrowski, X. Xiong, Z. Bukowski, R. Dybzinski, P.W. Klamut, J.E. Siewenie, O. Chmaissem, J. Shaffer, C.W. Kimball, J.D. Jorgensen, and S. Short, Phys. Rev. B - Condens. Matter Mater. Phys. **60**, 7006 (1999).

[6] J. Hemberger, A. Krimmel, T. Kurz, H.A. Krug von Nidda, V.Y. Ivanov, A.A. Mukhin, A.M. Balbashov, and A. Loidl, Phys. Rev. B - Condens. Matter Mater. Phys. **66**, 1 (2002).

[7] Y. Ogimoto, M. Izumi, A. Sawa, T. Manako, H. Sato, H. Akoh, M. Kawasaki, and Y. Tokura, Japanese J. Appl. Physics, Part 2 Lett. **42**, 0 (2003).

[8] Y. Yin, Y. Yin, X. Jiang, M.A. Koten, J.E. Shield, X. Chen, Y. Yun, A.T. N'Diaye, X. Hong, and X. Xu, Phys. Rev. Appl. **13**, 1 (2020).

[9] F.Y. Bruno, M.N. Grisolia, C. Visani, S. Valencia, M. Varela, R. Abrudan, J. Tornos, A. Rivera-Calzada, A.A. Ünal, S.J. Pennycook, Z. Sefrioui, C. Leon, J.E. Villegas, J. Santamaria, A. Barthélémy, and M. Bibes, Nat. Commun. **6**, 6306 (2015).

[10] O. Rousseau, S. Flament, B. Guillet, M.L.C. Sing, and L. Méchin, Proceedings **1**, 635 (2017).

[11] D. Pesquera, E. Khestanova, M. Ghidini, S. Zhang, A.P. Rooney, F. Maccherozzi, P. Riego, S. Farokhipoor, J. Kim, X. Moya, M.E. Vickers, N.A. Stelmashenko, S.J. Haigh, S.S. Dhesi, and N.D. Mathur, Nat. Commun. **11**, 3190 (2020).

[12] S.M. Wu, S.A. Cybart, P. Yu, M.D. Rossell, J.X. Zhang, R. Ramesh, and R.C. Dynes, Nat. Mater. **9**, 756 (2010).

[13] A. Quindeau, I. Fina, X. Marti, G. Apachitei, P. Ferrer, C. Nicklin, E. Pippel, D. Hesse, and M. Alexe, Sci. Rep. **5**, 1 (2015).

[14] A. Molinari, H. Hahn, and R. Kruk, Adv. Mater. **31**, (2019).

[15] A. Molinari, H. Hahn, and R. Kruk, Adv. Mater. **30**, 1703908 (2018).

[16] D. Yi, P. Yu, Y.C. Chen, H.H. Lee, Q. He, Y.H. Chu, and R. Ramesh, Adv. Mater. **31**, 1 (2019).

[17] C. Jin, Y. Zhu, W. Han, Q. Liu, S. Hu, Y. Ji, Z. Xu, S. Hu, M. Ye, and L. Chen, Appl. Phys. Lett. **117**, 252902 (2020).

[18] N. Zurauskiene, R. Lukose, S. Balevicius, V. Stankevic, S. Kersulis, V. Plausinaitiene, M. Vagner, and R. Navickas, IEEE Magn. Lett. **10**, 1 (2019).

[19] C. Moreno, C. Munuera, X. Obradors, and C. Ocal, Beilstein J. Nanotechnol. **3**, 722 (2012).

[20] A. Molinari, R. Witte, K.K. Neelisetty, S. Gorji, C. Kübel, I. Münch, F. Wöhler, L. Hahn, S. Hengsbach, K. Bade, H. Hahn, and R. Kruk, Adv. Mater. **32**, 1907541 (2020).

[21] L. Wang, C. Yang, J. Wen, S. Gai, and Y. Peng, J. Mater. Sci. Mater. Electron. **26**, 4618 (2015).

[22] H. Boschker, M. Huijben, a Vailionis, J. Verbeeck, S. van Aert, M. Luysberg, S. Bals, G. van Tendeloo, E.P. Houwman, G. Koster, D.H. a Blank, and G. Rijnders, J. Phys. D. Appl. Phys. **44**, 205001 (2011).

[23] P.M. Leufke, A.K. Mishra, A. Beck, D. Wang, C. Kübel, R. Kruk, and H. Hahn, Thin Solid Films **520**, 5521 (2012).

[24] D. Rasic, R. Sachan, N.K. Temizer, J. Prater, and J. Narayan, ACS Appl. Mater. Interfaces **10**, 21001 (2018).

[25] U.K. Sinha, B. Das, and P. Padhan, Nanoscale Adv. **2**, 2792 (2020).

[26] L. Cao, O. Petracic, P. Zakalek, A. Weber, U. Rücker, J. Schubert, A. Koutsioubas, S. Mattauch, and T. Brückel, Adv. Mater. **31**, 1 (2019).

[27] B. Paudel, B. Zhang, Y. Sharma, K.T. Kang, H. Nakotte, H. Wang, and A. Chen, Appl. Phys. Lett. **117**, (2020).

[28] S.K. Chaluvadi, F. Ajejas, P. Orgiani, S. Lebargy, A. Minj, S. Flament, J. Camarero, P. Perna, and L. Méchin, J. Phys. D. Appl. Phys. **53**, (2020).

[29] P. Zhou, Y. Qi, C. Yang, Z. Mei, A. Ye, K. Liang, Z. Ma, Z. Xia, and T. Zhang, AIP Adv. **6**, (2016).

[30] K. Wang, M.H. Tang, Y. Xiong, G. Li, Y.G. Xiao, W.L. Zhang, Z.P. Wang, Z. Li, and J. He, RSC Adv. **7**, 31327 (2017).

[31] S. Kumari, N. Mottaghi, C.Y. Huang, R. Trappen, G. Bhandari, S. Yousefi, G. Cabrera, M.S. Seehra, and M.B. Holcomb, Sci. Rep. **10**, 1 (2020).

[32] J.M.D. Coey, M. Viret, and S. von Molnár, Adv. Phys. **58**, 571 (2009).

[33] a. M. Haghiri-Gosnet, J. Wolfman, B. Mercey, C. Simon, P. Lecoeur, M. Korzenski, M. Hervieu, R. Desfeux, and G. Baldinozzi, J. Appl. Phys. **88**, 4257 (2000).

[34] Y. Takamura, R. V. Chopdekar, E. Arenholz, and Y. Suzuki, Appl. Phys. Lett. **92**, 162504 (2008).

[35] M. Mathews, E.P. Houwman, H. Boschker, G. Rijnders, and D.H. a. Blank, J. Appl. Phys. **107**, 013904 (2010).

[36] X. Wang, C. Jin, P. Wang, X. Pang, W. Zheng, D. Zheng, Z. Li, R. Zheng, and H. Bai, Appl. Phys. Lett. **115**, (2019).

[37] R. Herger, P.R. Willmott, C.M. Schlepütz, M. Björck, S. a. Pauli, D. Martoccia, B.D. Patterson, D. Kumah, R. Clarke, Y. Yacoby, and M. Döbeli, Phys. Rev. B - Condens. Matter Mater. Phys. **77**, 1 (2008).

[38] M. Huijben, L.W. Martin, Y.-H. Chu, M.B. Holcomb, P. Yu, G. Rijnders, D.H. a. Blank, and R. Ramesh, Phys. Rev. B **78**, 22 (2008).

[39] P.M. Leufke, R. Kruk, R. a. Brand, and H. Hahn, Phys. Rev. B **87**, 094416 (2013).

[40] M. Angeloni, G. Balestrino, N.G. Boggio, P.G. Medaglia, P. Orgiani, and A. Tebano, J. Appl. Phys. **96**, 6387 (2004).

[41] P.M. Leufke, R. Kruk, D. Wang, C. Kübel, and H. Hahn, AIP Adv. **2**, 032184 (2012).

[42] S. Valencia, a. Crassous, L. Bocher, V. Garcia, X. Moya, R.O. Cherifi, C. Deranlot, K. Bouzehouane, S. Fusil, a. Zobelli, a. Gloter, N.D. Mathur, a. Gaupp, R. Abrudan, F. Radu, a.

Barthélémy, and M. Bibes, Nat. Mater. **10**, 753 (2011).

[43] C. Thiele, K. Dörr, O. Bilani, J. Rödel, and L. Schultz, Phys. Rev. B - Condens. Matter Mater. Phys. **75**, 054408 (2007).

[44] A. Gambardella, P. Graziosi, I. Bergenti, M. Prezioso, D. Pullini, S. Milita, F. Biscarini, and V.A. Dediu, Sci. Rep. **4**, 1 (2014).

[45] R. Trappen, A.J. Grutter, C.Y. Huang, A. Penn, N. Mottaghi, S. Yousefi, A. Haertter, S. Kumari, J. Lebeau, B.J. Kirby, and M.B. Holcomb, J. Appl. Phys. **126**, (2019).

[46] L.W. Martin, Y.-H. Chu, and R. Ramesh, Mater. Sci. Eng. R Reports **68**, 89 (2010).

[47] M. Brahlek, A. Sen Gupta, J. Lapano, J. Roth, H. Zhang, L. Zhang, R. Haislmaier, and R. Engel-Herbert, Adv. Funct. Mater. **28**, 1702772 (2018).

[48] A. Molinari, P.M. Leufke, C. Reitz, S. Dasgupta, R. Witte, R. Kruk, and H. Hahn, Nat. Commun. **8**, 15339 (2017).

[49] T. Petrisor, M.S. Gabor, A. Boulle, C. Bellouard, C. Tiusan, O. Pana, and T. Petrisor, J. Appl. Phys. **109**, 123913 (2011).

[50] C. Aruta, G. Balestrino, a. Tebano, G. Ghiringhelli, and N.B. Brookes, Eur. Lett. **80**, 37003 (2007).

[51] S. Majumdar, H. Huhtinen, H.S. Majumdar, R. Laiho, and R. Österbacka, J. Appl. Phys. **104**, 033910 (2008).

[52] T.T. Fister, D.D. Fong, J. a. Eastman, P.M. Baldo, M.J. Highland, P.H. Fuoss, K.R.

Balasubramaniam, J.C. Meador, and P. a. Salvador, Appl. Phys. Lett. **93**, 151904 (2008).

[53] P.C. Bowes, J.N. Baker, J.S. Harris, B.D. Behrhorst, and D.L. Irving, Appl. Phys. Lett. **112**, 022902 (2018).

[54] T. Shi, Y. Chen, and X. Guo, Prog. Mater. Sci. **80**, 77 (2016).

[55] M. Siebenhofer, F. Baiutti, J. de Dios Sirvent, T.M. Huber, A. Viernstein, S. Smetaczek, C. Herzig, M.O. Liedke, M. Butterling, A. Wagner, E. Hirschmann, A. Limbeck, A. Tarancon, J. Fleig, and M. Kubicek, J. Eur. Ceram. Soc. (2021).

[56] K. Szot, W. Speier, U. Breuer, R. Meyer, J. Szade, and R. Waser, Surf. Sci. **460**, 112 (2000).

[57] K. Szot and W. Speier, Phys. Rev. B - Condens. Matter Mater. Phys. **60**, 5909 (1999).

[58] Y. Liang and D.A. Bonnell, Surf. Sci. **310**, 128 (1994).

[59] L. Brey, Phys. Rev. B - Condens. Matter Mater. Phys. **75**, 8 (2007).

[60] M. Bibes, L. Balcells, S. Valencia, J. Fontcuberta, M. Wojcik, E. Jedryka, and S. Nadolski, Phys. Rev. Lett. **87**, 67210 (2001).

[61] J. Simon, T. Walther, W. Mader, J. Klein, D. Reisinger, L. Alff, and R. Gross, Appl. Phys. Lett. **84**, 3882 (2004).

[62] J.Z. Sun, D.W. Abraham, R. a. Rao, and C.B. Eom, Appl. Phys. Lett. **74**, 3017 (1999).

[63] L. Li, Z. Liao, Z. Diao, R. Jin, E.W. Plummer, J. Guo, and J. Zhang, Phys. Rev. Mater. **1**, 034405 (2017).

[64] A. Tebano, A. Orsini, P.G. Medaglia, D. Di Castro, G. Balestrino, B. Freelon, A. Bostwick, Y.J. Chang, G. Gaines, E. Rotenberg, and N.L. Saini, Phys. Rev. B - Condens. Matter Mater. Phys. **82**, 1 (2010).

[65] a. Tebano, C. Aruta, S. Sanna, P.G. Medaglia, G. Balestrino, a. a. Sidorenko, R. De Renzi, G. Ghiringhelli, L. Braicovich, V. Bisogni, and N.B. Brookes, Phys. Rev. Lett. **100**, 2 (2008).

[66] D.A. Mota, Y. Romaguera Barcelay, A.M.R. Senos, C.M. Fernandes, P.B. Tavares, I.T. Gomes, P. Sá, L. Fernandes, B.G. Almeida, F. Figueiras, P. Mirzadeh Vaghefi, V.S. Amaral, A. Almeida, J. Pérez De La Cruz, and J. Agostinho Moreira, J. Phys. D. Appl. Phys. **47**, (2014).

[67] T. Becker, C. Streng, Y. Luo, V. Moshnyaga, B. Damaschke, N. Shannon, and K. Samwer, Phys. Rev. Lett. **89**, 21 (2002).

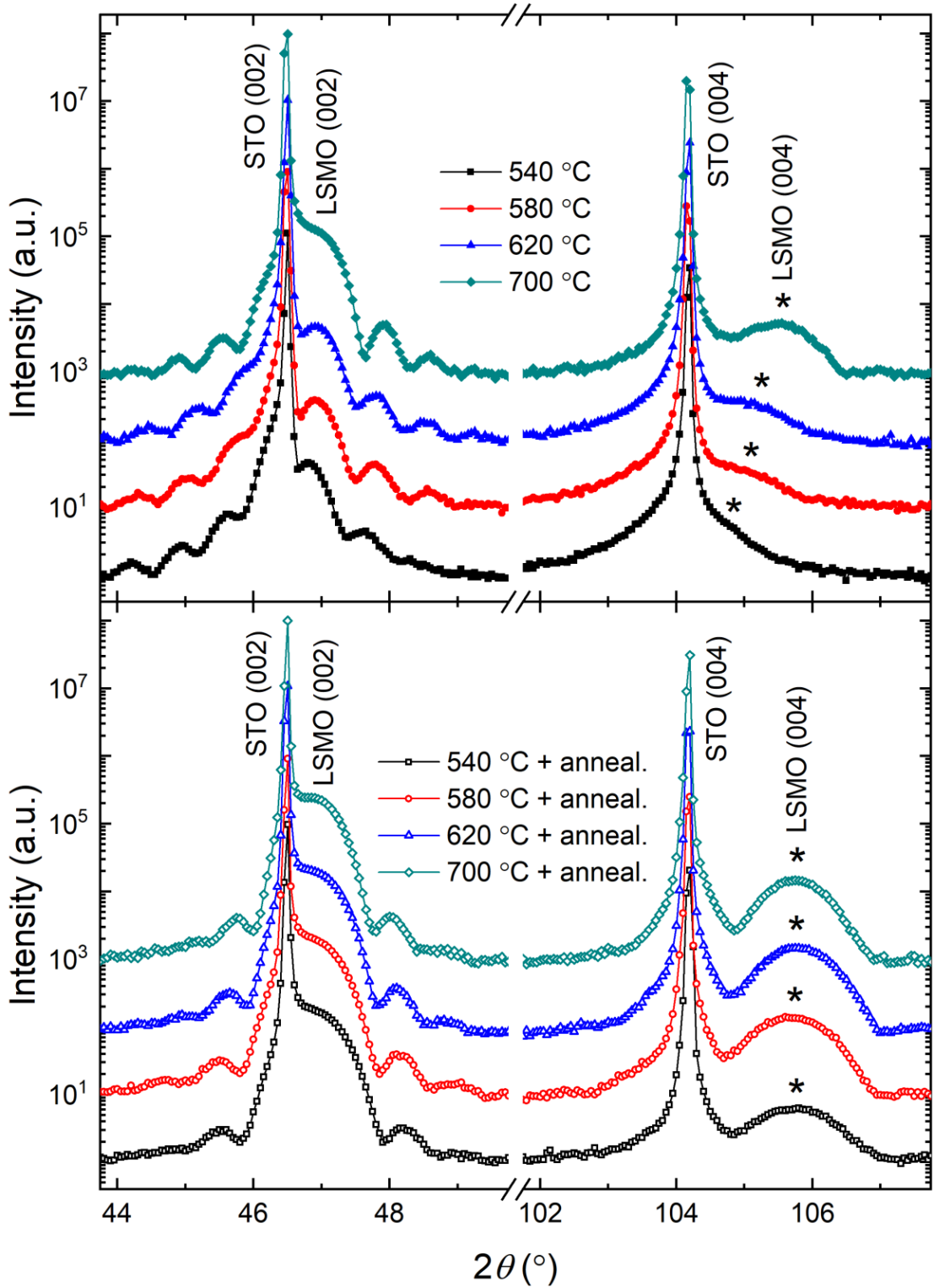

**Figure 1:** HRXRD scans of LSMO films with a thickness of about 13 nm deposited onto <001>-oriented STO substrate at various growth temperatures (top) and after post-deposition annealing (bottom). The presence of Laue fringes in the LSMO (002) reflections indicates a high-quality coherent stacking of LSMO unit cells on STO. The systematic shift towards higher angles of the LSMO reflections at higher growth temperatures (top) or after post-deposition annealing (bottom), as indicated by the asterisk symbol, is attributed to a reduction of the amount of oxygen deficiencies.

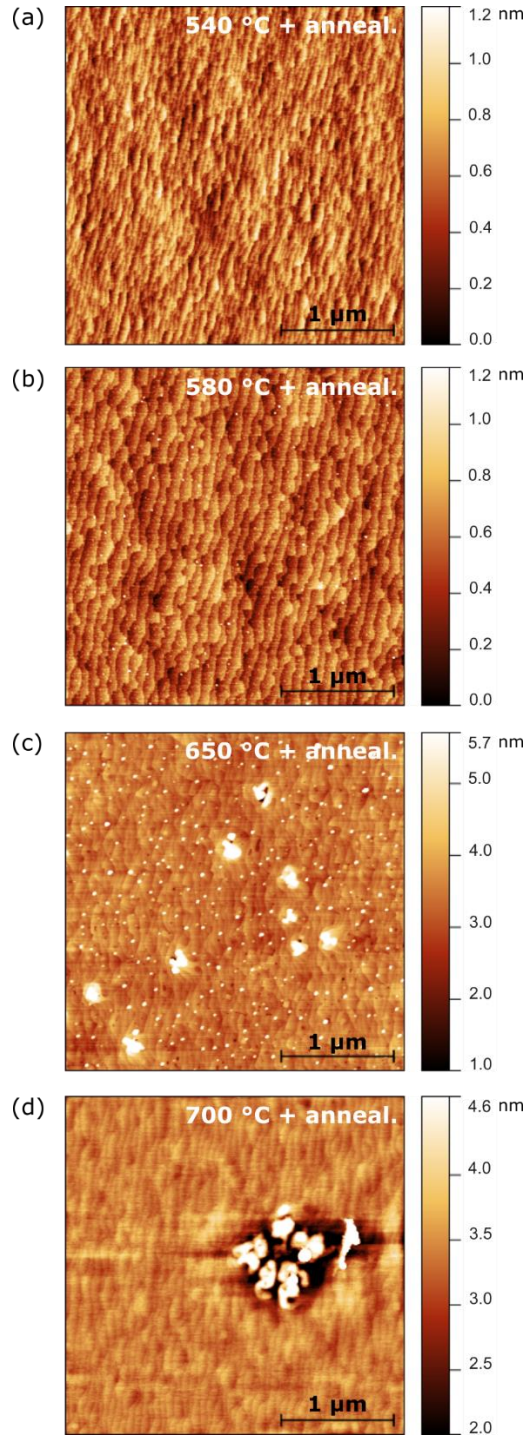

**Figure 2:** AFM images of the LSMO films grown at various temperatures after carrying out post-deposition annealing treatment. The surface morphology presents atomically flat terraces over large areas. Islands are formed on the LSMO surface at temperatures beyond 600 °C.

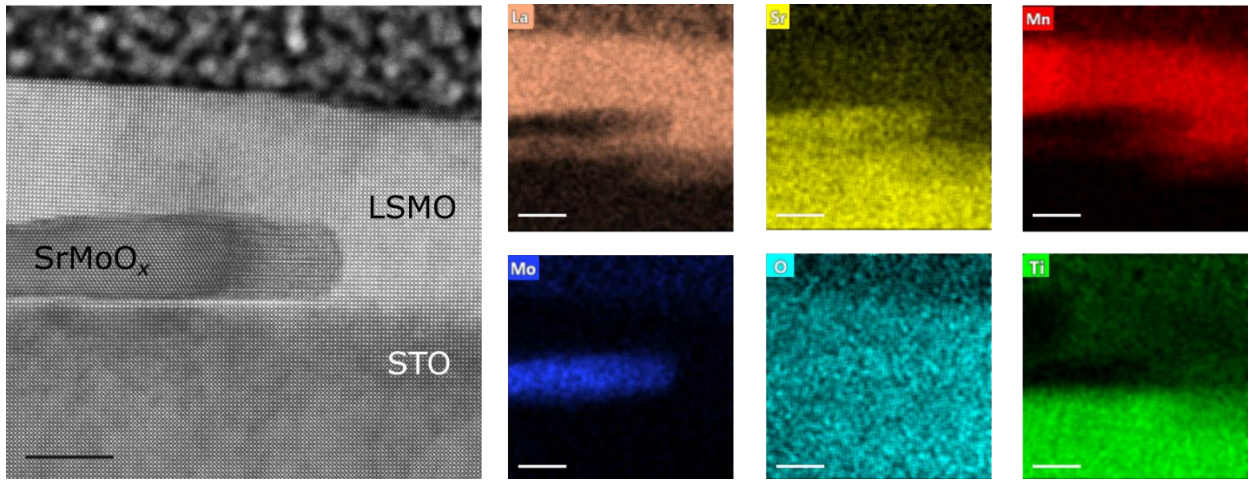

**Figure 3:** STEM-EDX characterization of a representative LSMO/STO cross section of a sample grown at 700 ˚C. A SrMoO$_x$ grain is present in the proximity of the film/substrate interface. The interfacial nature of the spurious grain indicates that its formation is related to the diffusion of Mo- and Sr- impurities from the bulk of the STO substrate. Despite the interfacial grain, LSMO epitaxy with cube-on-cube arrangement of the LSMO unit cells is preserved in the whole investigated area. The scale bar corresponds to 10 nm.

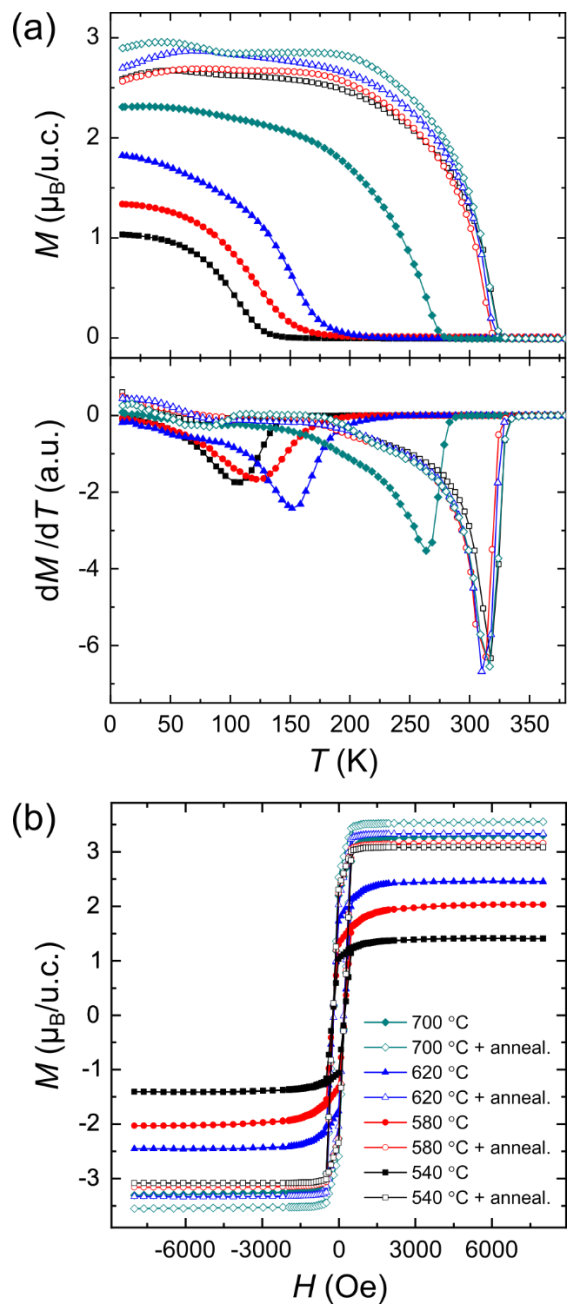

**Figure 4:** Magnetization and respective derivative curves as a function of temperature of LSMO films grown at different growth temperatures (filled symbols) and after post-deposition annealing (empty symbols) under application of a magnetic field of 100 Oe parallel to the in-plane film direction (a). Magnetic hysteresis loops carried out at 10 K (b). The systematic increase in magnetic moment and Curie temperature upon increasing the growth temperature, or after carrying out post-deposition annealing, is attributed to an improvement of the oxygen stoichiometry in LSMO.

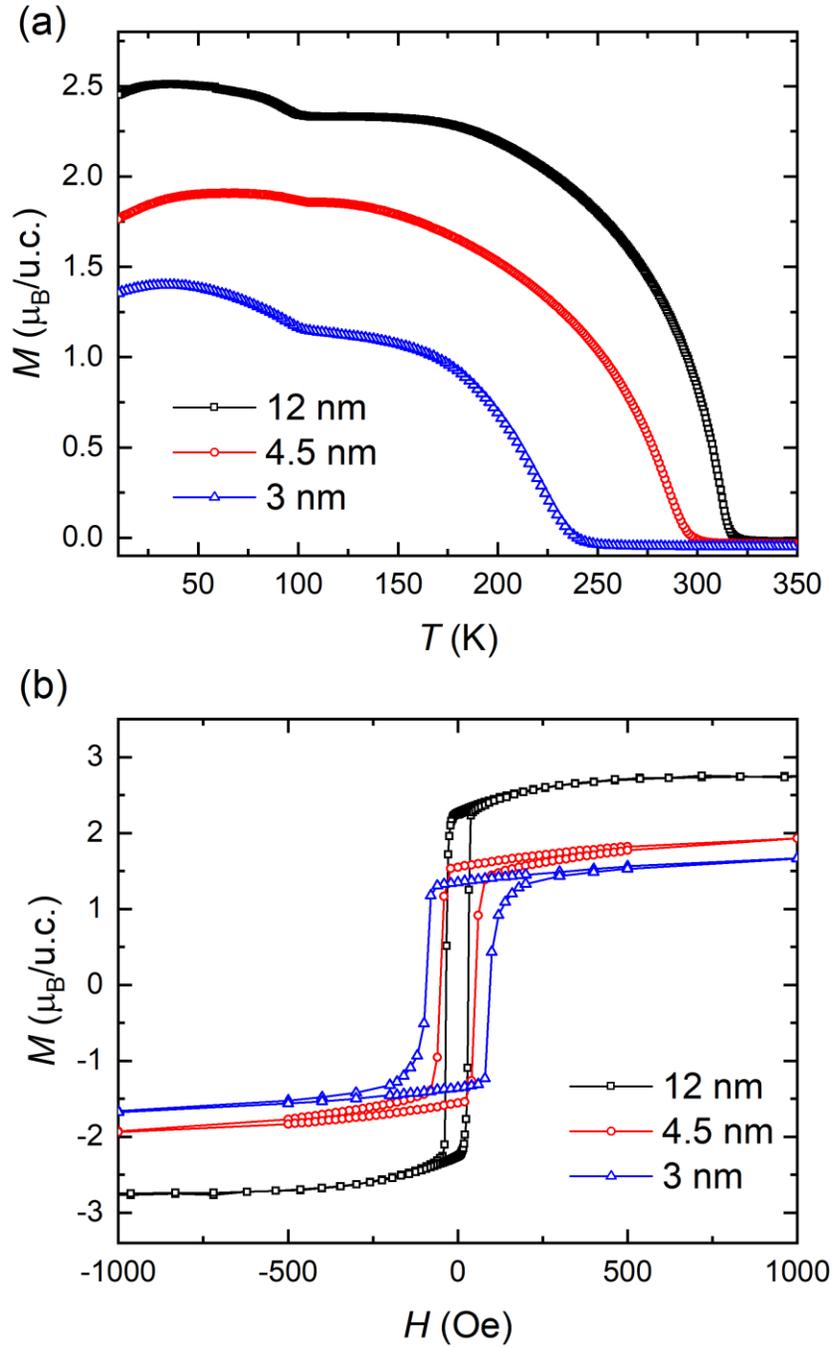

**Figure 5:** Magnetic field-cooled curves (a) and hysteresis loops at 10 K (b) of LSMO films with various thicknesses. The increase in magnetic moment of LSMO observable in the field-cooled curves at about 100 K is triggered by the structural phase transition of the STO substrate. Upon reducing the film thickness, LSMO reveals a systematic reduction of Curie temperature and magnetic moment as well as an increase in coercivity.

# SUPPLEMENTARY INFORMATION

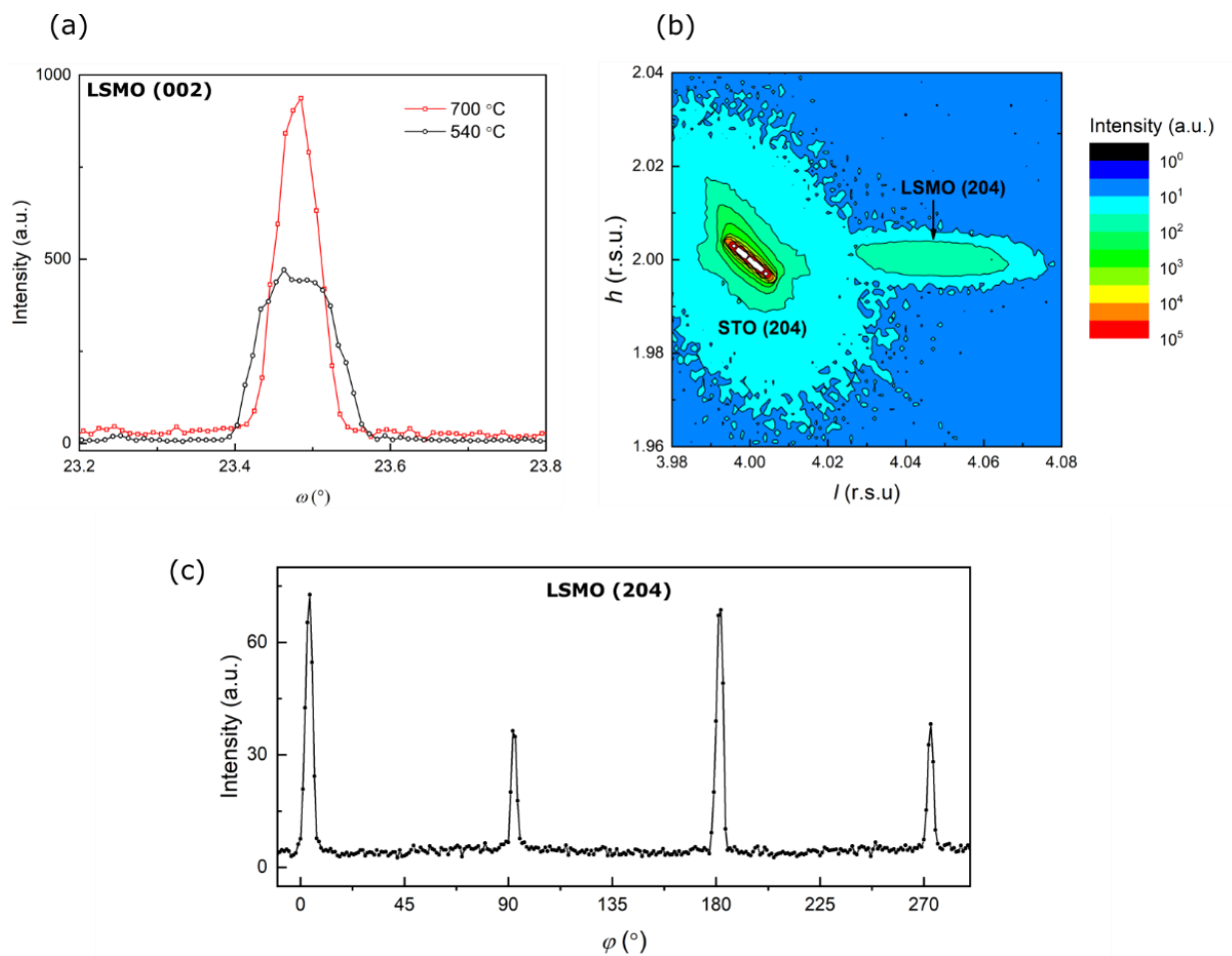

**Supplementary Figure 1**: Rocking curves of LSMO (002) reflections for annealed films grown at 540 and 700 °C (a). The broader rocking curve attained in the LSMO films grown at lower temperature suggests a slightly larger misalignment of the LSMO crystalline planes with respect to the STO substrate orientation. Two-dimensional reciprocal space map of the asymmetric (204) XRD reflection of an annealed LSMO film grown at 540 °C (b). LSMO grows fully-strained on STO, with an in-plane lattice parameter of about 3.90 Å. $\varphi$-scan of the asymmetric (204) reflection (c). The repeated occurrence of LSMO reflections every 90° is consistent with four-fold symmetry of the tetragonal unit cell of LSMO.

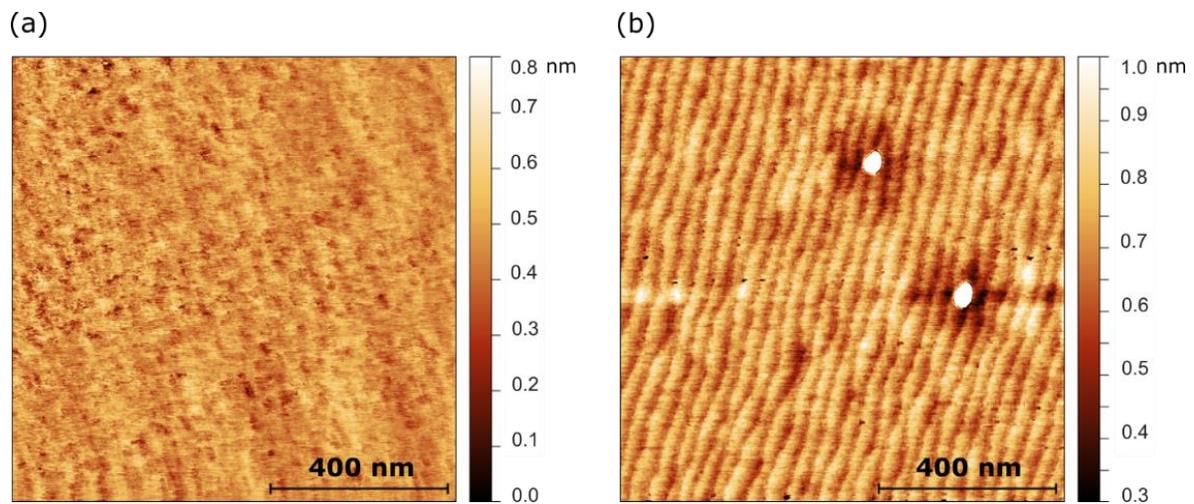

**Supplementary Figure 2:** Atomic force micrographs of (a) as-prepared ($T_{gr} \approx 620$ °C) and post-annealed LSMO films. The formation of atomically-flat terraces after carrying out annealing reveals the occurrence of LSMO surface reconstruction.

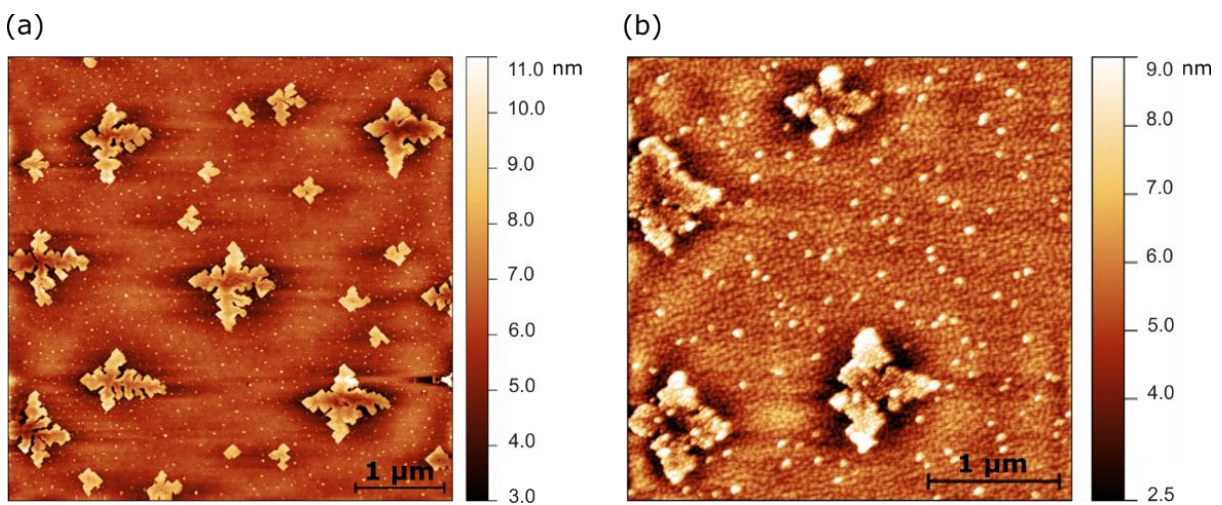

**Supplementary Figure 3:** (a) Atomic force micrograph of a STO substrate after exposure to a temperature of 700 °C inside the sputtering chamber (no LSMO deposition). (b) Surface morphology of a LSMO sample grown at 700 °C. The presence of dendritic-like islands in both cases evidences the fact that a high temperature promotes the diffusion of impurities from bulk STO.

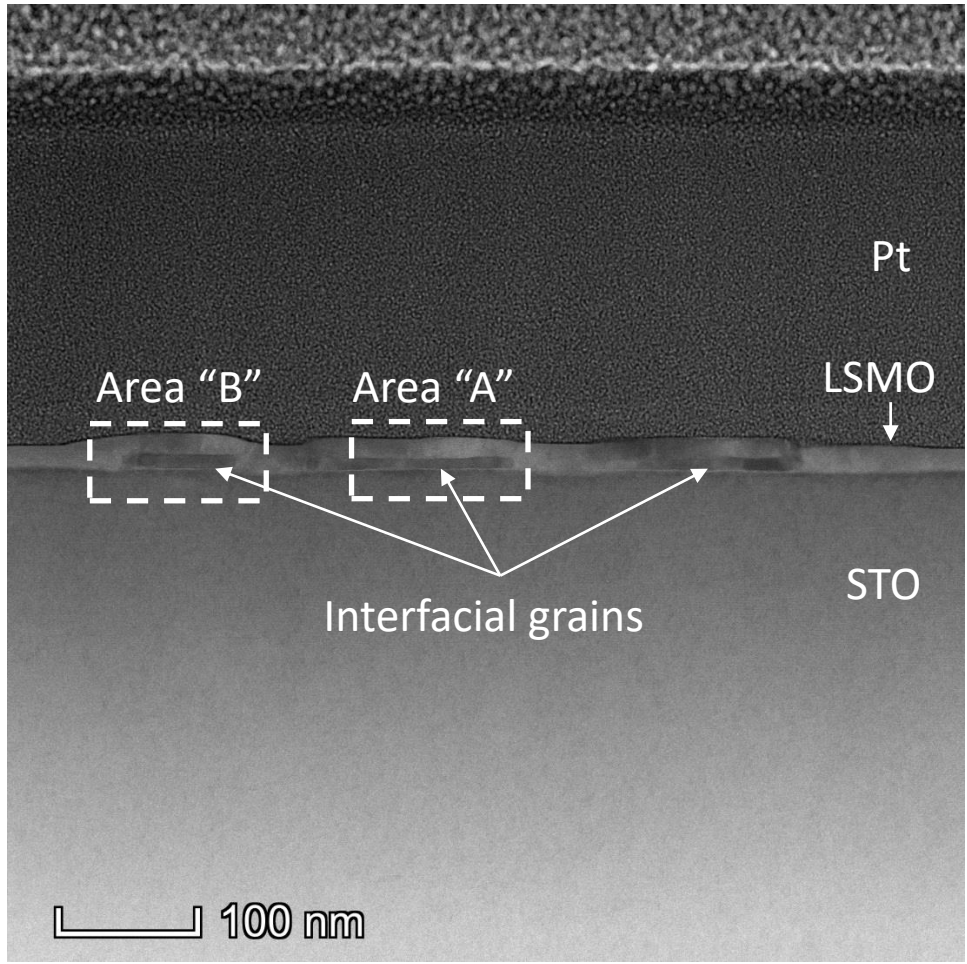

**Supplementary Figure 4:** Low-magnification STEM image showing a cross section of a LSMO/STO heterostructure grown at 700 °C. Several grains are present at the film/substrate interface. Further analysis of Area "A" and Area "B" is provided in the following.

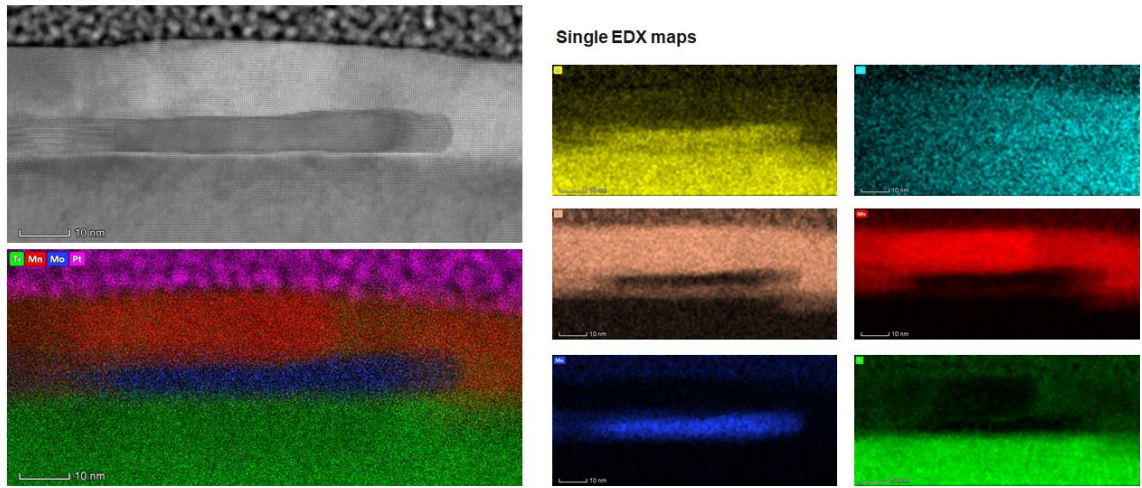

**Supplementary Figure 5:** STEM and EDX analyses of Area "A". The EDX maps reveal that the grain at the film/substrate interface is composed of a SrMoO$_x$ phase. The Sr/Mo ratio is close to 1-to-1 and constant within the particle.

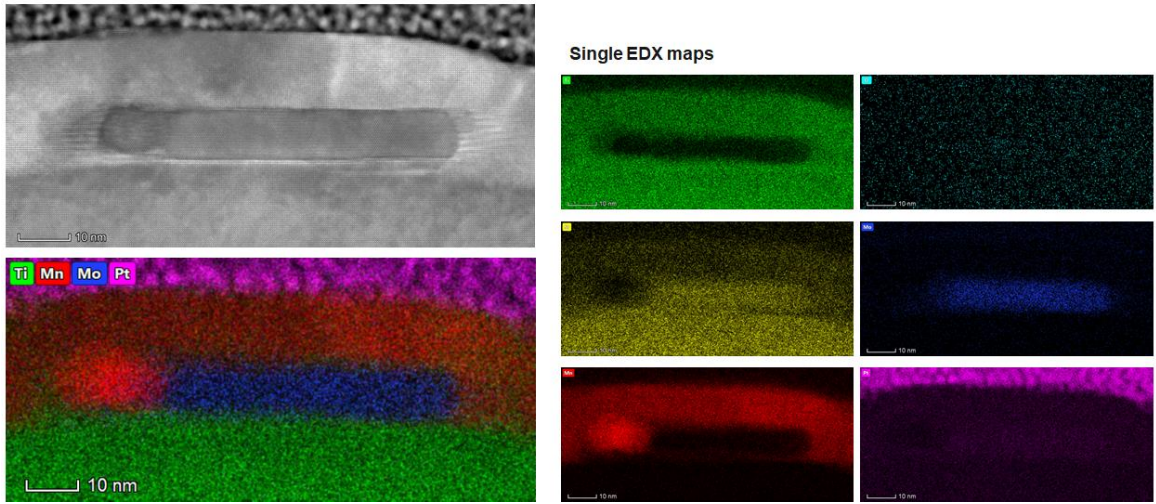

**Supplementary Figure 6:** STEM and EDX analyses of Area "B". In this case the interfacial grain is characterized by two distinguished phases: a Mn-rich phase (left side) and a SrMoO$_x$ phase (right side). This indicates that additional mechanisms of grain nucleation and growth, rather than just diffusion of bulk impurities, shall occur in LSMO/STO heterostructures.

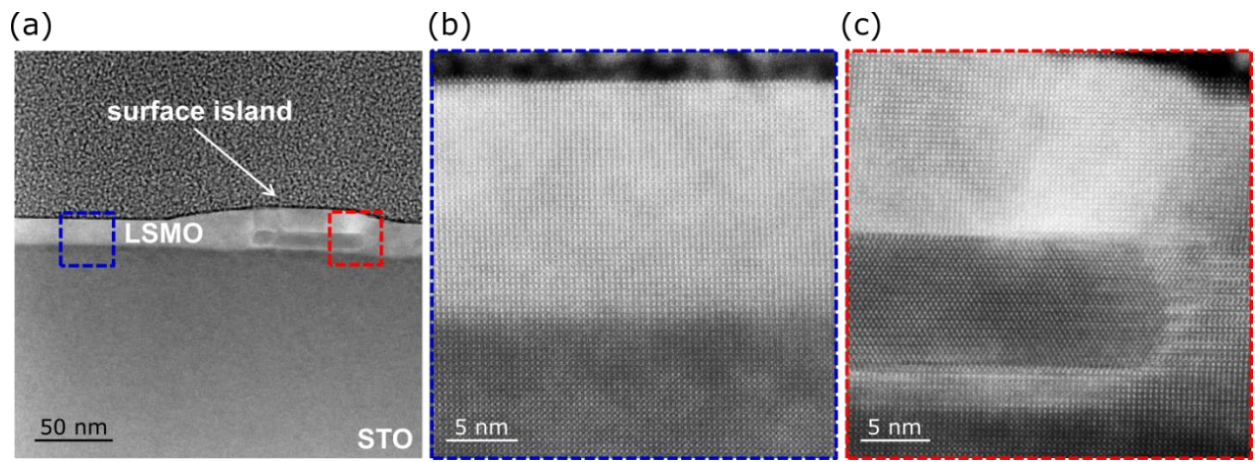

**Supplementary Figure 7:** STEM image close to "Area B" (a). Region with cube-on-cube arrangement of the LSMO unit cells onto STO and a sharp film/substrate interface (b). Region in proximity of a surface island with a different phase formed at the LSMO/STO interface; notably LSMO epitaxy is still preserved (c).

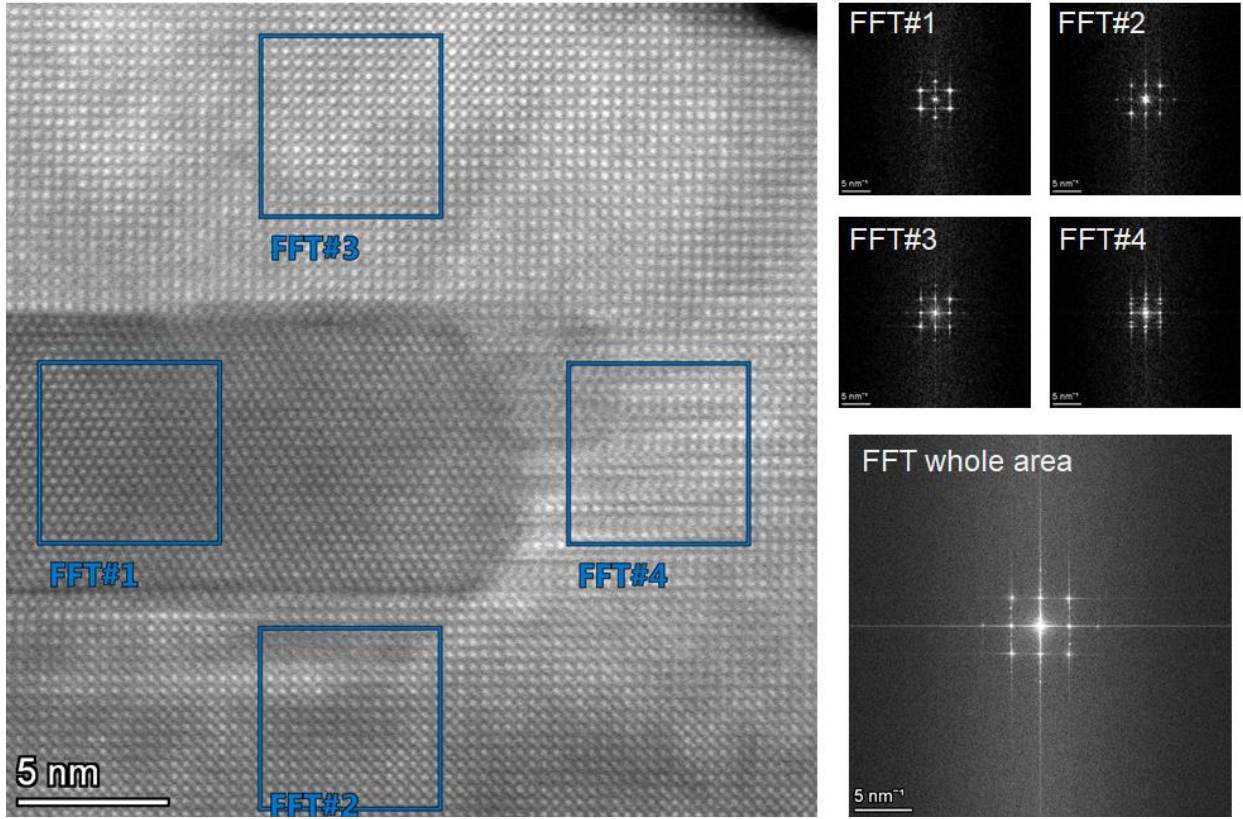

**Supplementary Figure 8:** High-magnification STEM image and Fast Fourier Transform (FFT) analysis of a LSMO/STO cross section in proximity of an interfacial grain. The nearly perfect match between FFT#2 and FFT#3 demonstrates that the LSMO film retains the epitaxial relation with the STO substrate, despite the presence of an interfacial grain with a different lattice orientation (FFT#1). The satellite spots observed in FFT#4 are possibly related to Moire' fringes.

**Treatment of epi-polished STO substrates**

The following steps describe the procedure for the preparation of $TiO_2$ single-terminated STO substrates, based on the work of Kareev et al., *Appl. Phys. Lett.* 93 (2008):

1. Ultrasonic cleaning of the original epi-polished STO substrate in deionized water at 70 °C for 25 min. The purpose is the formation of SrOH complexes on the STO surface.

2. Ultrasonic cleaning in HCl:HNO3 (3:1) at room temperature for 12 min. The acid selectively etches and removes the SrOH present at the STO surface, leaving a TiO2 terminated surface.

3. Rinsing of the STO substrates in DI water, followed by ultrasonic cleaning for 5 min in acetone and then 5 min in methanol at room temperature. These steps are necessary for removing any residuals of acidic contamination.

4. Annealing for 30 min at 1000 °C in air. This is fundamental in order to achieve atomically flat terraces with 1 u.c. steps (chemical etching alone is generally not enough).